\documentclass[submission,copyright,creativecommons]{eptcs}
\providecommand{\event}{TTC 2011} 
\usepackage[pdftex]{graphicx}
\usepackage{url}

\title{Hello World!\\An Instructive Case for the Transformation Tool Contest}
\author{Steffen Mazanek
\email{steffen.mazanek@gmail.com}
}
\def\authorrunning{Steffen Mazanek}
\def\titlerunning{Hello World Case}
\begin{document}
\maketitle

\begin{abstract}
This case comprises several primitive tasks that can be solved straight away with most transformation tools. The aim is to cover the most important kinds of primitive operations on models, i.e. create, read, update and delete (CRUD). To this end, tasks such as a constant transformation, a model-to-text transformation, a very basic migration transformation or diverse simple queries or in-place operations on graphs have to be solved. 

The motivation for this case is that the results expectedly will be very instructive for beginners. Also, it is really hard to compare transformation languages along complex cases, because the complexity of the respective case might hide the basic language concepts and constructs. 
\end{abstract}

\section{Introduction}
The case described in the following comprises not just a single transformation task. Rather it consists of a set of primitive tasks each of which can be solved with just a few lines of code with most transformation tools. So, this case is not a big challenge. However, it will result in an extensive set of small transformation programs that will be very instructive for beginners. Note that the goal of the Transformation Tool Contest (TTC) is to compare transformation languages/tools. However, the effort for investigating and comparing solutions for a complex case is remarkable. Moreover, it is really hard to learn the language constructs and concepts from large programs. Finally, it can be expected that solutions in many different transformation languages will be developed for this case, either for TTC 2011 or at a later time, because it is neither time consuming nor difficult to solve but still insightful. Actually, no developer of a transformation tool could effort not to solve the ``Hello World'' case.

\section{Tasks}
In the following the primitive tasks comprised by this case are introduced. The rationale of this selection is that the four basic functions create, read/query, update and delete (CRUD) should be covered. Note that certain subtasks are marked as optional, i.e. those are not required to solve the case but are considered only as extensions.

\subsection{Hello World!}

\begin{figure}
\centering
\includegraphics[width=.3\textwidth]{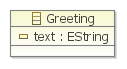}
\hspace{2cm}
\includegraphics[width=.25\textwidth]{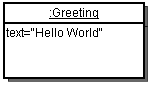}
\caption{The ``Hello World'' metamodel and the example instance}
\label{fig:helloworld}
\end{figure}

\begin{figure}
\centering
\includegraphics[width=.52\textwidth]{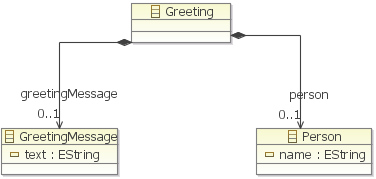}
\includegraphics[width=.43\textwidth]{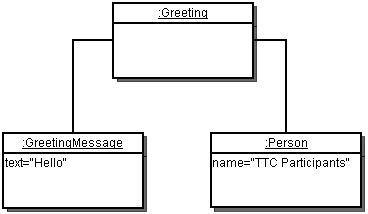}
\caption{The extended ``Hello World'' metamodel and the example instance}
\label{fig:helloworldext}
\end{figure}

\begin{itemize}
\item Provide a \emph{constant transformation} that creates the example instance of the ``Hello World'' metamodel given in Fig.~\ref{fig:helloworld}.
\item Consider now the slightly extended metamodel given in Fig.~\ref{fig:helloworldext}. Provide a \emph{constant transformation} that creates the \emph{model with references} as it is also shown in Fig.~\ref{fig:helloworldext}.
\item Next, provide a \emph{model-to-text transformation} that outputs the GreetingMessage of a Greeting together with the name of the Person to be greeted. For instance, the model given in Fig.~\ref{fig:helloworldext} should be transformed into the String \texttt{"Hello TTC Participants!"}.\footnote{Note that we provide as accompanying material on the case website a metamodel, \texttt{Result.ecore}, that contains classes for returning primitive results such as strings or numbers.}
\end{itemize}

\subsection{Count Matches with certain Properties}

\begin{figure}
\centering
\includegraphics[width=.6\textwidth]{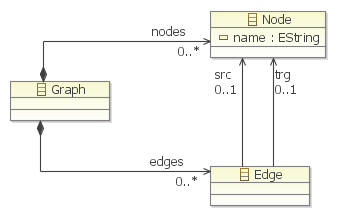}
\caption{The simple graph metamodel}
\label{fig:simplegraph}
\end{figure}

For the following \emph{querying} tasks the input models are simple graphs conforming to the metamodel given in Fig.~\ref{fig:simplegraph}. As results numbers should be returned, again wrapped into an object of the respective result type.

\begin{figure}
\centering
\includegraphics[width=.5\textwidth]{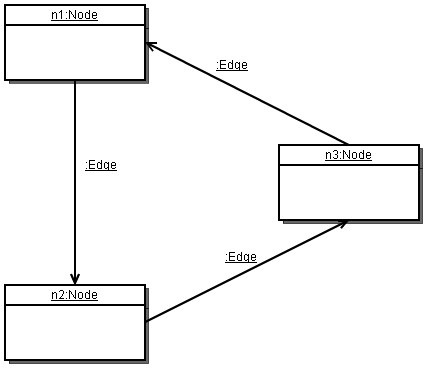}
\caption{Circle of three nodes (simplified representation of edge objects)}
\label{fig:circle}
\end{figure}

\begin{itemize}
\item Provide a model query that counts the number of nodes in a graph.
\item Provide a model query that counts the number of looping edges in a graph, i.e. edges where the source and the target node coincide.
\item Provide a model query that counts the number of isolated nodes in a graph, i.e. nodes that are neither the source nor the target of any edge.
\item Provide a model query that counts the number of matches of a circle consisting of three nodes, i.e. the pattern shown in Fig.~\ref{fig:circle} where n1, n2 and n3 are pairwise distinct. Note that each circle in the model should be matched three times.
\item Optional: Provide a model query that counts the number of dangling edges in a graph, i.e. edges where either the source or the target node is missing.
\end{itemize}

\subsection{Reverse Edges}
For the remainder of the paper we assume that there are no dangling edges in our graphs, i.e. src and trg of edges are properly set.
Provide a transformation that reverses all edges in a graph conforming to the simple graph metamodel given in Fig.~\ref{fig:simplegraph}. This is an \emph{update} operation.

\subsection{Simple Migration}
\begin{figure}
\centering
\includegraphics[width=.9\textwidth]{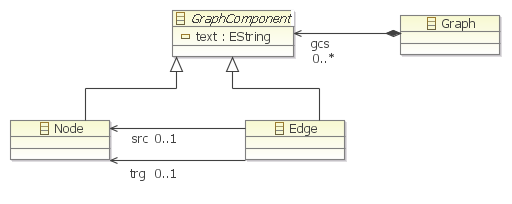}
\caption{The evolved graph metamodel}
\label{fig:graph2}
\end{figure}

Provide a transformation that migrates a graph conforming to the metamodel given in Fig.~\ref{fig:simplegraph} to a graph conforming to the metamodel given in Fig.~\ref{fig:graph2}. The \texttt{name} of a node becomes its \texttt{text}. The \texttt{text} of a migrated edge has to be set to the empty string.

\begin{figure}
\centering
\includegraphics[width=.4\textwidth]{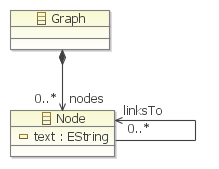}
\caption{The even more evolved graph metamodel}
\label{fig:graph3}
\end{figure}

Optional: Provide a topology-changing migration that transforms graphs of the metamodel given in Fig.~\ref{fig:simplegraph} to graphs as defined by the metamodel in Fig.~\ref{fig:graph3}.

\subsection{Delete Node with Specific Name and its Incident Edges}

Given a simple graph conforming to the metamodel of Fig.~\ref{fig:simplegraph}. Provide a transformation that \emph{deletes} the node with name ``n1''. If a node with name ``n1'' does not exist, nothing needs to be changed. It can be assumed that there is at most one occurrence of a node with name ``n1''.

Optional: Provide a transformation that removes the node ``n1'' (as above) and all its incident edges.

\subsection{Optional: Insert Transitive Edges}
\begin{figure}
\centering
\includegraphics[width=.4\textwidth]{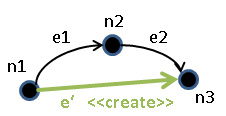}
\caption{Insertion of transitive edges}
\label{fig:trans}
\end{figure}

Consider the input graph as a relation $R$. Provide a transformation that creates the graph corresponding to the relation $R\cup R^2$. To this end, for every three nodes n1, n2 and n3 and two edges e1, e2 where e1 points from n1 to n2 and e2 points from n2 to n3, insert an additional edge pointing from n1 to n3, if there is no edge connecting n1 and n3 already (cf.\ Fig.~\ref{fig:trans}).

\section{Evaluation}
Since this case is easy to solve an official award was not appropriate here. Instead, every team that presented a correct solution for all subtasks at TTC 2011 has received "Hello World" TTC cups for all team members. In addition, an evaluation sheet was filled in by the participants. The following criteria have been used: completeness, understandability and conciseness (all with equal weights).

\section{Conclusion}
This case comprises several easily solvable tasks. The results can still be expected to be very useful. Most importantly, they can be explored by beginners in order to get an impression how basic problems can be solved using the different transformation approaches. Moreover, it is planned to use these tasks and the corresponding solutions as an initial fill for the recently developed online judge for model transformations (\url{http://transformation-judge.org}), a system that supports the upload of cases and solutions, which will be black-box-tested and automatically ranked according to criteria such as performance, LOC and user ratings.

The website of this case proposal is \url{http://sites.google.com/site/helloworldcase/}. All the metamodels as well as example input/output models can be downloaded from this site. In addition, this site contains all solutions presented at TTC 2011.

\textbf{Acknowledgment:} I would like to thank Pieter Van Gorp for proposing a meaningful rationale for the actual task selection and for his idea to call this case ``Hello World''.


\end{document}